\documentclass[12pt]{article}
\usepackage{latexsym,amssymb,amsmath}
\usepackage{amsthm, amstext}
\usepackage{array, amsfonts, mathrsfs}
\usepackage[dvips]{graphicx}

\usepackage{epsf}

\newcommand{\cho}{{\rm ch}}
\newcommand{\shi}{{\rm sh}}
\newcommand{\tha}{{\rm th}}

\numberwithin{equation}{section}

\begin{document}
\date{}

\author{M.I.Belishev\thanks {Saint-Petersburg Department of the Steklov Mathematical
Institute, Russian Academy of Sciences; Saint-Petersburg State
University; belishev@pdmi.ras.ru},\,\,A.V.Ivanov\thanks
{Saint-Petersburg State University;
                 regul1@mail.ru.}}
\title{On a calculus of variations problem}

\maketitle

\begin{abstract}
The paper is of scientific-methodical character. The classical
soap film shape (minimal surface) problem is considered, the film
being stretched between two parallel coaxial rings. An analytical
approach based on relations to the Sturm-Liouville problem is
proposed. An energy terms interpretation of the classical
Goldschmidt condition is discussed. Appearance of the soliton
potential in course of the second variation analysis is noticed.
\medskip

\smallskip


\end{abstract}

\section{Introduction}
\subsubsection*{Setup}
In the space endowed with the standard Cartesian coordinate system
$x,y,z$, there are two rings $\{(x,y,z)\,|\,\,x=\mp h,\,\,
y^2+z^2=1\}\,.$ Between the rings, a soap film is stretched, the
film minimizing its area owing to the surface stretch forces. By
symmetry of the physical conditions, the film takes the shape of a
rotation (around $x$-axis) surface, whereas to find this shape is
to solve the well-known minimization problem for the functional
 \begin{equation}\label{basic functional}
S_h[y]:=2\pi \int_{-h}^h y(x)\sqrt{1+{y'}^2(x)}\, dx
 \end{equation}
provided the boundary conditions
\begin{equation}\label{boundary conditions}
y(-h)=y(h)=1\,.
 \end{equation}
The value $h>0$, which is equal to the half-distance between the
rings, plays the role of the basic parameter. The goal of the
paper is to study the behavior of the solutions to the problem
(\ref{basic functional}), (\ref{boundary conditions}) depending on
$h$.

\subsubsection*{Results}
In the above-mentioned or analogous setup, the given problem is
considered (at least on a formal level) in almost all manuals on
the calculus of variations. It is studied in detail in the
monograph \cite{Busl}, whereas we deal with the version of the
manual \cite{Arf}, which will be commented on later. In our paper:
\smallskip

\noindent$\bullet$\,\,\,a purely analytical way of solving the
problem \footnote{\cite{Busl} provides the treatment in
geometrical terms of the extremals behavior: see pages 28--45},
which uses the well-known facts of the Sturm-Liouville theory, is
proposed
\smallskip

\noindent$\bullet$\,\,\,the case of the critical value $h=h_*$
(such that the problem turns out to be unsolvable for $h>h_*$) is
studied in detail, the study invoking the third variation of the
functional $S_{h_*}[y]$
\smallskip

\noindent$\bullet$\,\,\,a criticism of the arguments of \cite{Arf}
concerning to the Goldschmidt condition, is provided; our own
interpretation of the lack of solvability for $h>h_*$ based on the
energy considerations is proposed.
\smallskip

\noindent A noteworthy point is that, in course of studying the
second variation of the functional (\ref{basic functional}), the
key role is played by the Sturm-Liouville equation with 1-soliton
potential. However, we didn't succeed in finding a satisfactory
explanation for this fact.

\subsubsection*{Acknowledgements}
The work is supported by the grants RFBR 11-01-00407A and SPbSU\\
6.38.670.2013. The authors thank A.F.Vakuenko for the useful
discussions and consultations.

\section{Extremum investigation}

\subsubsection*{Extremals}
Let us recall the well-known facts. The extremals of the
functional (\ref{basic functional}) satisfy the Euler equation
$$F_y-\frac{d}{dx}F_{y'}\,=\,0\,,$$
where $F(y,y'):=y\sqrt{1+{y'}^2}$. It possesses the first integral
$F-y'F_{y'}= C$; the consequent integration provides the solutions
of the form $y(x,C_1,C_2)=C_1{}\cho\frac{x+C_2}{C_1}$. The
conditions (\ref{boundary conditions}) easily imply $C_2=0$, which
leads to the 1-parameter family of the extremals
 \begin{equation}\label{extr general}
y(x,C)\,=\,C \cho\frac{x}{C}\,, \qquad C>0\,.
 \end{equation}
The functional value at an extremal is found by integration:
 \begin{equation}\label{S on extr general}
S_h[y( \cdot ,C)]\,=\,2\pi \int_{-h}^hC
\cho\frac{x}{C}\,\sqrt{1+\shi^2\frac{x}{C}}\,dx=2\pi Ch+\pi C^2
\shi\frac{2h}{C}\,.
 \end{equation}

\subsubsection*{Solvability conditions}
Substituting $x=h$ to (\ref{extr general}) with regard to
(\ref{boundary conditions}), one gets the equation $C
\cho\frac{h}{C}=1$ for determination of the constant $C$, which
can be written in the form
 \begin{equation}\label{transcend eq}
\phi(\tau)\,=\,\frac{1}{h}\,, \qquad \text{where} \quad
\tau:=\frac{h}{C}>0\,, \,\, \phi(\tau):=\frac{\cho \tau}{\tau}\,.
 \end{equation}
Elementary analysis provides the following facts.
\smallskip

\noindent$\bullet$\,\,\, The function $\phi$ is downward convex,
whereas $\phi(\tau)\to \infty$ holds for $\tau \to 0$ and $\tau
\to \infty$. It has only one positive minimum at the point
$\tau=\tau_*$ determined by the equality $\phi^\prime(\tau)=0$.
The latter is equivalent to a transcendent equation
 \begin{equation}\label{eq for tau*}
1-\tau \tha \tau \,=\,0\,.
 \end{equation}
 \begin{figure}[tbp]
\centering
\includegraphics[width=3in,height=2in]{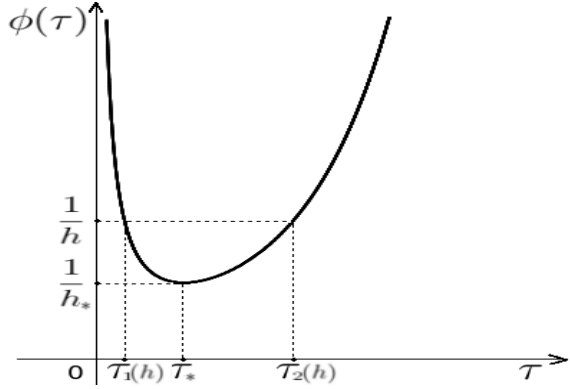}
\caption{Function $\phi$} \label{fig1}
 \end{figure}

\noindent$\bullet$\,\,\,The equation (\ref{transcend eq}) is
solvable if $h \leqslant h_*$ only, where
$h_*:=\frac{1}{\phi(\tau_*)}$. For $h<h_*$ it has two distinct
roots $\tau_{1,2}(h):\,\tau_{1}(h)<\tau_{2}(h)$; for $h=h_*$ the
roots coincide. For $h \to 0$ one has $\tau_1(h) \to 0$ and
$\tau_2(h) \to \infty$, the relations
 \begin{equation}\label{asympt tau}
\lim \limits_{h\to 0}\frac{\tau_1(h)}{h}\,=\,1\,, \qquad \lim
\limits_{h\to 0}h\,\frac{e^{\tau_2(h)}}{\tau_2(h)}\,=\,2
 \end{equation}
being valid.
\smallskip

\noindent$\bullet$\,\,\, The function $\tau_1(h)$ defined for
$0\leqslant h \leqslant h_*$ is invertible; the inverse function
is
 $$h\,=\,h(\tau_1)\,\overset{(\ref{transcend
eq})}=\,\frac{\tau_1}{\cho \tau_1}\,, \qquad 0\leqslant
\tau_1\leqslant \tau_*\,.$$ For the latter, we have
$$\frac{dh}{d\tau_1}\,=\,
\frac{1-\tau_1 \tha \tau_1}{\cho\tau_1}$$ that leads to
\begin{equation}\label{dtau/dh=infty}
\frac{d\tau_1}{dh}\,=\, \frac{\cho\tau_1}{1-\tau_1 \tha
\tau_1}\,,\qquad \lim \limits_{h \to
h_*}\frac{d\tau_1}{dh}\,\overset{(\ref{eq for tau*})}=\,\infty\,.
 \end{equation}
\smallskip

\begin{figure}[tbp]
\centering
\includegraphics[width=5in,height=1.5in]{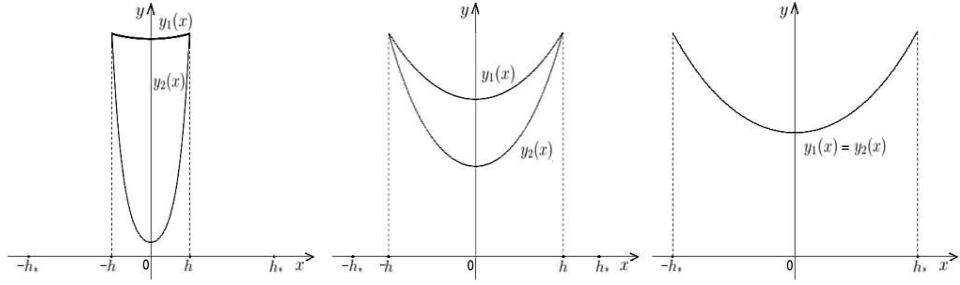}
\caption{The extremals} \label{fig2}
 \end{figure}

In particular, the aforesaid shows that for $h \leqslant h_*$ the
functional $S_h[y]$ possesses two extremals
 \begin{equation}\label{extremals y 1,2}
y_{1,2}(x)\,\overset{(\ref{extr
general})}=\,\frac{h}{\tau_{1,2}(h)}\,\cho\left[\frac{\tau_{1,2}(h)}{h}\,x\right]\,,
\qquad -h \leqslant x \leqslant h\,,
 \end{equation}
which are distinct if $h<h_*$ and coincide if $h=h_*$. {\it For
$h>h_*$, the functional $S_h[y]$ does not possess extremals}. On
fig \ref{fig2}, the graphs of extremals are shown for a small,
intermediate, and critical values of the distance $h$.

Substituting (\ref{extremals y 1,2}) to (\ref{extr general}), one
gets the equalities
 \begin{equation}\label{S[y 1,2]}
S_h\left[y_{1,2}\right]\,=\,2\pi\frac{h^2}{\tau_{1,2}(h)}+\pi\,\frac{h^2}{\tau_{1,2}^2(h)}\,\shi\,2
\tau_{1,2}(h)\,, \qquad 0<h\leqslant h_* \,.
 \end{equation}
By them, with regard to (\ref{asympt tau}), one easily derives the
relations
 \begin{equation*}
\lim \limits_{h\to 0}S_h[y_1]\,=\,0\,, \qquad \lim \limits_{h\to
0}S_h[y_2]\,=\,2\pi\,.
 \end{equation*}
Some additional analysis implies
 \begin{equation}\label{S[y 1]<S[y_2]}
S_h[y_1]\,<\,S_h[y_2]\,, \qquad 0<h<h_*,
 \end{equation}
wheraes $S_{h_*}[y_1]=S_{h_*}[y_2]$ holds by coincidence of the
extremals for $h=h_*$ (see fig \ref{fig3}).

\subsubsection*{Second variation}
Testing the extremals on the presence of extremum, we use the
Taylor representation
 \begin{equation}\label{Taylor repres S}
 S_{h}[y+t\eta]\,\underset{t\,\sim\, 0}=\,S_{h}[y]+t \delta S_{h}[y;h]+t^2 \delta^2 S_{h}[y;h] + t^3 \delta^3 S_{h}[y;h] + o(t^3)
 \end{equation}
\footnote{here, for the extremals $y$, one has $\delta
S_{h}[y;h]=0$} and, in particular, the second variation. Its
general form at the extremals (\ref{extr general}) is derived by
the straightforward differentiation:
 \begin{equation*}
\delta^2S_h[y;
\eta]:=\frac{1}{2!}\,\left[\frac{d^2}{dt^2}\,S_h[y+t\eta]\right]\bigg|_{t=0}\overset{(\ref{basic
functional}), (\ref{extr general})}=\frac{1}{2C} \int_{-h}^h
\frac{C^2 {\eta'}^2(x)-\eta^2(x)}{\cho^2\frac{x}{C}}\,dx\,,
 \end{equation*}
where $\eta \in C^1[-h,h]$ is a test function,
$\eta(-h)=\eta(h)=0$. Introducing a new variable $s=\frac{x}{C}$
and test function
 \begin{equation}\label{test function}
\psi(s):=\frac{\eta(Cs)}{\cho s} \qquad (\,\text{so\,
that}\,\,\,\,\eta(x)=\psi\left(\frac{x}{C}\right)\cho\frac{x}{C}\,)\,,
 \end{equation}
and integrating by parts with regard to
$\psi(-\tau)=\psi(\tau)=0$, after some simple calculation we get
 \begin{equation}\label{delta^2S}
\delta^2S_h[y; \eta]=\alpha \int_{-\tau}^\tau \left[{\psi'}^2(s)
-\frac{2}{\cho^2s}\,\psi^2(s)\right]\,ds\,,
 \end{equation}
where $\alpha={\rm const}>0$.

Let us consider the integral in (\ref{delta^2S}) as a functional
of $\psi$. For it, the corresponding Euler equation takes the form
of the Sturm-Liouville equation
 \begin{equation}\label{soliton eq}
\psi^{\prime \prime}+\frac{2}{\cho^2s}\psi\,=\,0
 \end{equation}
with the {\it soliton potential} $q=\frac{2}{\cho^2s}$. We did not
succeed to recognize, whether it  appears in the given problem
just by occasion, or there is a deeper reason for that.

We study the second variation by the use of the special solution
to (\ref{soliton eq}) of the form
 \begin{equation}\label{soluton mu}
\mu(s)\,:=\,1-s \,\tha s\,.
 \end{equation}
It is distinguished by the conditions $\mu(0)=1$ and
$\mu(-s)=\mu(s)$, has the ordinary roots $\tau=\mp\tau_*$ (see
(\ref{eq for tau*})), and is {\it positive} in the interval
$(-\tau_*,\tau_*)$. Recall that, outside its roots, any solution
to the equation $u''+qu=0$ satisfies the well-known Riccati
equation
 $$
\left[\frac{u'}{u}\right]^\prime+\left[\frac{u'}{u}\right]^2=-q\,.
 $$
Applying this to the solution $\mu$, we have
 \begin{equation}\label{Riccati for mu}
\left[\frac{\mu'}{\mu}\right]^\prime+\left[\frac{\mu'}{\mu}\right]^2=-\frac{2}{\cho^2\tau}\,.
 \end{equation}
Therefore, for $|\tau| < \tau_*$ (outside the roots of $\mu$) the
following transformations of the integral in (\ref{delta^2S}) turn
out to be quite correct:
 \begin{align*}
& \int_{-\tau}^\tau \left[{\psi'}^2(s)
-\frac{2}{\cho^2s}\,\psi^2(s)\right]\,ds\overset{(\ref{Riccati for
mu})}=\int_{-\tau}^\tau \left\{ {\psi'}^2+
\left[\left(\frac{\mu'}{\mu}\right)^\prime+\left(\frac{\mu'}{\mu}\right)^2\right]\,\psi^2\right\}\,ds\overset{(*)}=\\
& =\int_{-\tau}^\tau \left[ {\psi'}^2-
\frac{\mu'}{\mu}\,\left(\psi^2\right)^\prime
+\left(\frac{\mu'}{\mu}\right)^2\,\psi^2\right]\,ds=\int_{-\tau}^\tau
\left[ {\psi'}- \frac{\mu'}{\mu}\,\psi\right]^2 ds\,.
 \end{align*}
Integrating by parts in the equality $(*)$, one uses the boundary
conditions $\psi(\mp \tau)=0$. The same conditions yield that
$\frac{\psi}{\mu}$ is bounded as $|s|\leqslant \tau_*$, what
enables one to justify the derivation also in the case
$\tau=\tau_*$.

As a consequence, for the second variation (\ref{delta^2S}) on any
test function $\eta$, we have:
 \begin{equation}\label{Repres delta^2 S}
\delta^2S_h[y; \eta]=\beta\,\int_{-\tau}^\tau \left[ {\psi'}-
\frac{\mu'}{\mu}\,\psi\right]^2 ds=
    \begin{cases}
>0 & \text{for}\,\,\,\tau<\tau_*\\\geqslant 0 & \text{for}\,\,\,\tau=\tau_*
    \end{cases}
    ,
 \end{equation}
the equality $\delta^2S_h[y; \eta]=0$ (for $\tau=\tau_*$) being
valid only on the function $\eta$, which corresponds  (in the
meaning of (\ref{test function})) to the function $\psi= c\mu$
with a constant $c\not=0$.

\subsubsection*{Extremal $y_1$}
Fix an  $h<h_*$; for it, the equality $\tau_1(h)<\tau_*$ holds. By
the latter, the representation (\ref{Repres delta^2 S}) is valid
with $\tau<\tau_*$ that implies  $$\delta^2S_h[y_1; \eta]\,>\,0$$
for any test function $\eta$. Consequently, on the extremal $y_1$
the functional $S_h[y]$ does attain a minimum. Its minimal value
is determined by (\ref{S[y 1,2]}). As is seen from (\ref{S[y
1]<S[y_2]}), this minimum is {\it local} (is not global).

\subsubsection*{Extremal $y_2$}
For  $h<h_*$, one has $\tau_2(h)>\tau_*$ and the representation
(\ref{Repres delta^2 S}) becomes invalid. Let us show that the
variation $\delta^2S_h[y_2; \eta]$ turns out to be sign-indefinite
and takes negative values on appropriate $\eta$.

Consider the boundary value spectral problem
\begin{align}
\label{string 1} & \psi^{\prime
\prime}+\lambda\frac{2}{\cho^2s}\psi\,=\,0\,, \qquad
-\tau<s<\tau\\
\label{string 2} & \psi(-\tau)\,=\,\psi(\tau)\,=\,0
 \end{align}
for an inhomogeneous string with the density
$\rho=\frac{2}{\cho^2s}$ and the fixed endpoints. Here $\tau$ is a
parameter. Recall the well-known facts (see, e.g.,
\cite{Atkinson}).
\smallskip

\noindent$\bullet$\,\,\,The problem possesses the ordinary
discrete spectrum $\{\lambda_k(\tau)\}_{k \geqslant 1}$:
$$0<\lambda_1(\tau)<\lambda_2(\tau)<\dots,\quad \lambda_k(\tau)\underset{k \to \infty}\to
\infty\,,$$ whereas the corresponding eigenfunctions
$\{\psi_k(\cdot\,; \tau)\}_{k \geqslant 1}$ constitute an
orthogonal basis of the space  $L_{2, \,\rho}(-\tau,\tau)$.
\smallskip

\noindent$\bullet$\,\,\,The relation
$\frac{d\lambda_k(\tau)}{d\tau}<0$ holds, so that the eigenvalues
are the strictly monotonic decreasing functions of $\tau$.
\smallskip

\noindent$\bullet$\,\,\,The first (minimal) eigenvalue is
 \begin{equation}\label{lambda 1 = min}
\lambda_1(\tau)\,=\,\min \limits_{0\not=\psi \in
H^1_0[-\tau,\tau]}\frac{ \int_{-\tau}^\tau {\psi'}^2(s)\,ds}{
\int_{-\tau}^\tau \frac{2}{\cho^2s}\,\psi^2(s)\,ds}\,,
 \end{equation}
where $H^1_0[-\tau,\tau]:=\{y\,|\,\,y,y'\in
L_2(-\tau,\tau),\,y(\mp\tau)=0\}$ is the Sobolev space. For $\tau
\to 0$ one has $\lambda_1(\tau)\to \infty$.
\smallskip

\noindent$\bullet$\,\,\,The eigenfunction $\psi_1$ has no roots in
$-\tau<s<\tau$. The functions $\psi_k$ of the numbers $k\geqslant
2$ {\it do have} the roots into this interval.
\smallskip

\noindent By the above mentioned facts, the behavior of the low
bound of the string spectrun is the following. For $\tau\sim 0$,
we have $\lambda_1(\tau)\gg 1$. As $\tau$ grows, the value of
$\lambda_1(\tau)$ is decreasing, whereas for $\tau=\tau_*$ one has
$\lambda_1(\tau_*)=1$ and $\psi_1 = c \mu$. Indeed, for
$\tau=\tau_*$ the equation (\ref{soliton eq}) \footnote{which is
the same as the equation (\ref{string 1}) with $\lambda=1$}
possesses the solution $\psi=\mu$, which satisfies the conditions
(\ref{string 2}), i.e., is an eigenfunction of the string
corresponding to $\lambda=1$. It is namely the {\it first}
eigenfunction since $\mu$ has no roots into $(-\tau_*,\tau_*)$.

Further, for $\tau>\tau_*$, by monotonicity of the eigenvalues, we
have $\lambda_1(\tau)<1$. Therefore, by (\ref{lambda 1 = min}),
there is a function $\psi_0 \in H^1_0[-\tau,\tau]$ satisfying
 $$
\frac{ \int_{-\tau}^\tau {{\psi_0}'}^2(s)\,ds}{ \int_{-\tau}^\tau
\frac{2}{\cho^2s}\,\psi_0^2(s)\,ds}\,<\,1
 $$
that is equivalent to
 $$
\int_{-\tau}^\tau \left[{{\psi_0}'}^2(s)
-\frac{2}{\cho^2s}\,{\psi_0}^2(s)\right]\,ds\,<\,0\,.
 $$
Therefore, for the function $\eta_0$ related with $\psi_0$ via the
relation (\ref{test function}), by virtue of (\ref{delta^2S}) one
has:
$$\delta^2S_h[y_2; \eta_0]\,<\,0\,.$$
Hence, the extremal $y_2$ provides no extremum to the functional
$S_h[y]$.

\subsubsection*{Critical case}
The previous considerations deal with the case $h<h_*$. Now, let
$h=h_*$, so that the extremals do coincide:
$$y_1(x)=y_2(x)\overset{(\ref{extremals y 1,2})}=\frac{h_*}{\tau_*}\,
\cho\left[\frac{\tau_*}{h_*}\,x\right]\,=:\,y_*(x)\,, \qquad -h_*
\leqslant x \leqslant h_*\,.$$ Let us show that there is no
extremum at $y_*$. Recall that the function $\mu$ is defined in
(\ref{soluton mu}).
\smallskip

Find the variations of the functional $S_{h_*}[y]$, choosing
$$\eta_*(x)\overset{(\ref{test function})}=\,\mu\left(\frac{\tau_*}{h_*}\,x\right)
\cho\left[\frac{\tau_*}{h_*}\,x\right]$$ as a test function. By
the choice, we have $\delta S_{h_*}[y_*;\eta_*]=0$ и $\delta^2
S_{h_*}[y_*;\eta_*]\overset{(\ref{Repres delta^2 S})}=0$. Let us
find the third variation. As one can easily verify, on the
arbitrary element and test function it is of the form
 \begin{align*}
& \delta^3 S_{h}[y;\eta]:=\frac{1}{3!}\,\left[\frac{d^3}{dt^3}\,S_h[y+t\eta]\right]\bigg|_{t=0}=\\
& =
\pi\int_{-h}^{h}\frac{{\eta'}^2(x)}{\left(1+{y'}^2(x)\right)^{\frac{3}{2}}}
\left[\eta(x)-\frac{y(x)y'(x)\eta'(x)}{1+{y'}^2(x)}\right] dx\,.
 \end{align*}
Taking $h=h_*,\,y=y_*$ и $\eta=\eta_*$, the simple calculation
provides:
 \begin{equation*}
\delta^3 S_{h_*}[y_*; \eta_*] = \frac{2\pi
\tau_*^4}{3h_*}\,\not=\,0\,.
 \end{equation*}
By (\ref{Taylor repres S}), we have
$S_{h_*}[y_*+th_*]\,\underset{t\,\sim\,0}=\,\gamma\,t^3 + o(t^3)$
с $\gamma \not=0$ that certifies the absence of extremum.

\section{Comments}

\subsubsection*{On the Goldschmidt condition}
For $h>h_*$, the functional (\ref{basic functional}) with the
conditions (\ref{boundary conditions}) does not have extremals at
all. In \cite{Arf} (chapter 17, section 2), this fact is
accomplished with the following qualitative explanation.

As $h$ grows, the area o the film is growing. For sufficiently big
$h$, by energy reasons, it turns out to be more profitable for the
film to fill the both of the rings separately and, so, take the
total area $\pi 1^2+\pi 1^2=2\pi$. By this, the film breaks,
whereas the critical area value turns out to be  $2\pi$, which is
declared as the {\it Goldschmidt break condition}.

\begin{figure}[tbp]
\centering
\includegraphics[width=3in,height=2in]{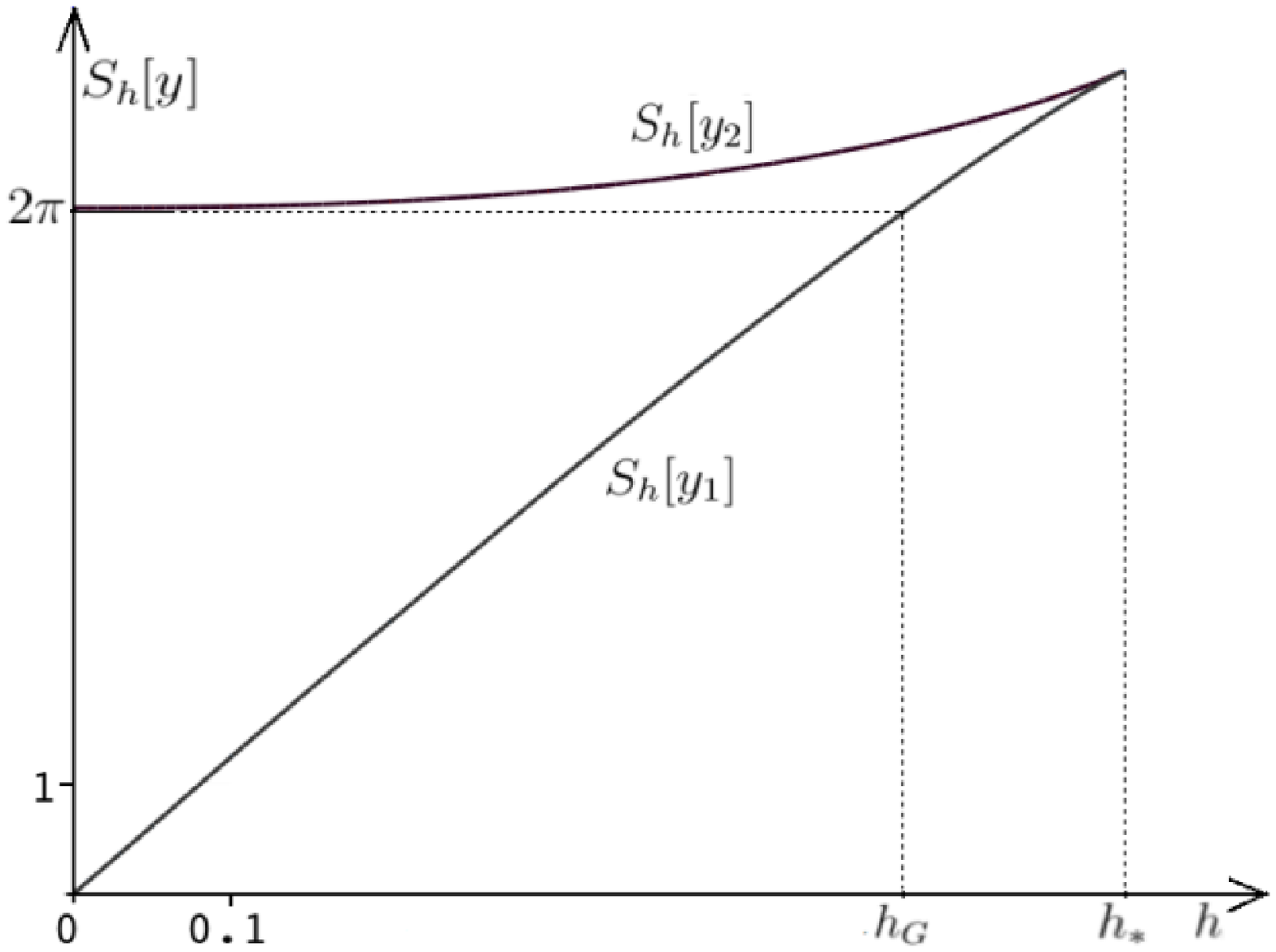}
\caption{Площади} \label{fig3}
 \end{figure}

The given explanation is incorrect \footnote{It is the matter,
which has inspired our interest to the problem}. Defining the
`Goldschmidt constant' $h_G$ as the solution of the equation
$S_h[y_1]=2\pi$ (with respect to $h$), it is easy to recognize
that it is solvable and
$$ 0.5277... = h_G<h_*=0.6627...\,,$$
so that the corresponding extremal $y_1$ does exist and describes
a stable pre-critical shape of the film: see fig \ref{fig3}.

Probably, the incorrect explanation is just a result of confusion.
In the first exercise at the end of the section (page 689), the
reader is proposed `to find such a value of $h$ that the rotation
surface area is equal to the total area of the end rings'.

\subsubsection*{Physical considerations}
Is the value $h=h_*$ distinguished from a physical viewpoint?
Bellow we propose a variant of the answer on this question.
\smallskip

Let $h<h_*$ and the film be of the shape described by the extremal
$y_1$. Contacting with the rings, the film influences on them by
the surface stretch forces, the rings being attracted with each
other. As the distance between them grows, the system accumulates
a potential energy. In the framework of the model under
consideration, one can assume the potential energy of the stretch
forces to be proportional to the film area $S_h[y_1]$. The
derivative
$$-\,\frac{dS_h[y_1]}{dh}=:F(h)$$
may be naturally \footnote{by analogy to the model of elastic
spring, where $E_{\rm pot}=\frac{kh^2}{2}$ and $F=-E_{\rm
pot}^\prime=-kh$} interpreted as a force of the rings attraction.
Let us find its value.

Denoting $R(\tau):= \frac{2}{\tau}+\frac{\shi 2\tau}{\tau^2}\,,$
we have $S_h[y_1]\,\overset{(\ref{S[y 1,2]})}=\,\pi h^2
R(\tau_1(h))$ that follows to $$ F(h)=-\,2\pi h R(\tau_1(h))-\pi
h^2 R^\prime(\tau_1(h))\tau_1^\prime(h)\,.$$ Implementing the
differentiation in the right hand side, after the simple
transformations with regard to the first of the equalities
(\ref{dtau/dh=infty}), we get
 $$F(h)\,=\,-\,4\pi\,\frac{h}{\tau_1(h)}\,.$$
Differentiating one more time, we arrive at the relations
$$F^\prime(h)\,=\,-\,4\pi\,\frac{\tau_1(h)-h\tau_1^\prime(h)}{\tau_1^2(h)}\,
\overset{(\ref{dtau/dh=infty})}{\quad\,\,\underset{h \to h_*}\to
\infty}\,.$$ Such a behavior motivates to regard the value $h=h_*$
as critical: one may assume that it is the infinite velocity of
the force growing, which leads to the break of the film, and
forbids its existence for  $h>h_*$.

\end{document}